\documentclass[%
reprint,
superscriptaddress,
 amsmath,amssymb,
prb,
]{revtex4-1}

\usepackage{graphicx}
\usepackage{dcolumn}
\usepackage{bm}
\usepackage{color}

\begin{document}

\preprint{APS/123-QED}

\title{
Phonon and Magnetic Excitations in Superconducting Ca$_{10}$Pt$_4$As$_8$(Fe$_{1-x}$Pt$_x$As)$_{10}$
}

\author{K. Ikeuchi}
 \email{k\_ikeuchi@cross.or.jp}
\affiliation{%
Research Center for Neutron Science and Technology,
Comprehensive Research Organization for Science and
Society (CROSS), Tokai, Ibaraki 319-1106, Japan
}%
\author{Y. Kobayashi}
\author{K. Suzuki}
\author{M. Itoh}
\affiliation{%
Department of Physics, Nagoya University, Nagoya 464-8602, Japan
}%
\author{R. Kajimoto}
\affiliation{
Materials and Life Science Division (MLF),
J-PARC Center, Tokai, Ibaraki 319-1195, Japan
}%
\author{P. Bourges}
\affiliation{%
Laboratoire L\'{e}on Brillouin, CEA-CNRS, CEA Saclay,
91191 Gif-sur-Yvette Cedex, France
}%
\author{A. D. Christianson}
\affiliation{%
Quantum Condensed Matter Division,
Oak Ridge National Laboratory, Oak Ridge, TN 37831, USA
}%
\author{H. Nakamura}
\author{M. Machida}
\affiliation{%
CCSE, Japan Atomic Energy Agency,
5-1-5 Kashiwanoha, Kashiwa, Chiba, 277-8587, Japan
}%
\author{M. Sato$^{1, }$}%
 \email{m\_sato@cross.or.jp}

\date{\today}

\begin{abstract}
Inelastic neutron scattering studies have been carried out on selected phonons and magnetic excitations of a crystal of superconducting (SC) Ca$_{10}$Pt$_4$As$_8$(Fe$_{1-x}$Pt$_x$As)$_{10}$ with the onset transition temperature  $T_{\rm c}^{\rm onset} \sim$ 33 K to investigate the role that orbital fluctuations play in the Cooper pairing.
The spectral weight of the magnetic excitations, $\chi ''({\bm Q}, \omega)$ at ${\bm Q} = {\bm Q}_{\rm M}$ (magnetic $\Gamma$ points) is suppressed (enhanced) in the relatively low (high) $\omega$ region.
The maximum of the enhancement is located at $\omega = \omega_{\rm p} \sim$ 18 meV at temperature $T = 3$ K corresponding to the $\omega_{\rm p}/k_{\rm B}T_{\rm c}^{\rm onset} \sim$ 6.3.
This large value is rather favorable to the orbital-fluctuation mechanism of the superconductivity with the so-called $S_{++}$ symmetry.
In the phonon measurements, we observed slight softening of the in-plane transverse acoustic mode corresponding to the elastic constant $C_{66}$.
This softening starts at $T$ well above the superconducting $T_{\rm c}$, as $T$ decreases.
An anomalously large increase in the phonon spectral weight of in-plane optical modes was observed in the range of $35 < \omega < 40$ meV with decreasing $T$ from far above $T_{\rm c}$.
Because this $\omega$ region mainly corresponds to the in-plane vibrations of Fe atoms, the finding presents information on the coupling between the orbital fluctuation of the Fe 3$d$ electrons and lattice system, useful for studying possible roles of the orbital fluctuation in the pairing mechanism and appearance of the so-called nematic phase. 
\end{abstract}

\pacs{Valid PACS appear here}
\maketitle


\section{introduction}
In the study of physics of Fe-based superconductors,\cite{01kamihara} it is important to understand what roles the orbital fluctuations plays in realizing the high superconducting (SC) transition temperature $T_{\rm c}$.
Of course, it is widely known that various characteristics of the systems, two dimensional layers of strongly correlated electrons and superconductivity in the neighbourhood of the magnetically ordered phase, suggest that the spin-fluctuation exchange is relevant to the SC pairing\cite{02mazin,03kuroki} and that the so-called $S_{\pm}$ symmetry of the order parameter $\Delta$ is realized with its sign reversal between the Fermi surfaces around $\Gamma$ and M points in the reciprocal space (${\bm Q}$ space).
However, the situation is not simple, because the possibility of the $S_{++}$ symmetry without the sign reversal was pointed out based on the observed small rates of the $T_{\rm c}$ suppression by non-magnetic impurities much smaller than those expected for the $S_{\pm}$ symmetry.\cite{04kawabata,05sato1,06kawamata,07sato2}
It has been also shown theoretically that the orbital fluctuation mechanism can explain the high-$T_{\rm c}$ superconductivity with the $S_{++}$ symmetry as well as various other experimental results which were simply regarded as evidences for the $S_{\pm}$ symmetry.\cite{08onari2,11onari3,10onari1,9kontani1,13kontani2,12onari4,14inoue,15kontani3}
Among these works, the primary roles of the orbital fluctuation enhanced by the Aslamazov-Larkin type vertex correction due to the spin fluctuation is emphasized.

Experimentally, the phase diagram of Ba(Fe$_{1-x}$Co$_x$)$_2$As$_2$ has attracted much attention,\cite{16ni1,17kim} because we can find many physical characteristics which are common to many other Fe-based systems.
One example is the existence of the tetragonal-orthorhombic structural transition at $T_{\rm S}$ higher than the antiferromagnetic transition temperature $T_{\rm N}$, implying that the orbital degrees of freedom should be taken into consideration besides the spin degrees of freedom.
Another example is that even in the macroscopically tetragonal phase, the breakdown of the four fold symmetry becomes appreciable with decreasing $T$, in static quantities such as the electrical resistivity\cite{18chu} and energy splitting of the electronic band formed by the $3d_{yz}$ and $3d_{zx}$ orbitals\cite{19yi,20shimojima} at around the characteristic temperature ($T^*$) significantly higher than $T_{\rm S}$ (we call $T^*$ nematic temperature here.).
In arguing these findings, it seems to be inevitable to have roles of orbital fluctuation in mind, as stressed by Zhang {\it et al}.\cite{021zhang} for example.
Now, to answer the question ``{\it spin or orbital?} " has become primarily important for the understanding of the physics of strongly correlated electrons entangled with spins, orbitals and probably lattice systems.
We have carried out neutron scattering studies on magnetic excitations and phonons of a Ca$_{10}$Pt$_4$As$_8$(Fe$_{1-x}$Pt$_x$As)$_{10}$ crystal with the onset SC transition temperature $T_{\rm c}^{\rm onset} \sim$ 33 K to find effects of the orbital fluctuation on these dynamical behaviors of Fe-based systems.
In the phonon measurements, we focused on particular phonons near the $\Gamma$ and M points in its ${\bm Q}$ space of the pseudo tetragonal cell.
These reciprocal points were chosen, because the orbital fluctuation is expected to be strong.\cite{9kontani1,13kontani2}
Parts of the results of studies have been published in Refs.~\onlinecite{22sato3,23ikeuchi,24sato4}.

\section{experiments and calculations}
\begin{figure}
\includegraphics[scale=0.38,angle=-90]{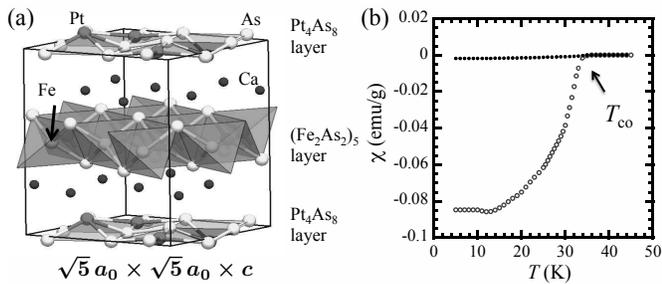}
\caption{\label{fig:xtal}
(a) Unit cell of Ca$_{10}$Pt$_4$As$_8$(Fe$_{1-x}$Pt$_x$As)$_{10}$ is shown schematically.
(b) Diamagnetic signals of a part of the sample.
The closed and open circles show the result in field and zero-field cooling conditions, respectively, where the field was applied parallel to the $c$ direction.
}
\end{figure}

Ca$_{10}$Pt$_4$As$_8$(Fe$_{1-x}$Pt$_x$As)$_{10}$ was discovered by Kakiya {\it et al}.\cite{21kakiya}
It has the stacking of --(Fe$_{1-x}$Pt$_x$As)--Ca--PtAs$_2$--Ca-- layer units as shown schematically in Fig.~\ref{fig:xtal}(a).
Within the $a$-$b$ planes, the so-called $\sqrt{5}\,a_0\times\sqrt{5}\,a_0$ unit cell is formed, $a_0$ ($ \sim$3.903 \AA) being the lattice parameter of pseudo tetragonal cell of the planes\cite{21kakiya,25ni2} having two Fe atoms in an FeAs plane.
The PtAs$_2$ planes were found, by NMR $1/T_1$,\cite{26kobayashi} to be essentially insulating and nonmagnetic, and although the crystal structure is rather complicated, we can study the physical properties of the FeAs conducting planes of Ca$_{10}$Pt$_4$As$_8$(Fe$_{1-x}$Pt$_x$As)$_{10}$, mainly considering the electronic structure similar to other Fe-based superconductors.\cite{27neupane,28hirupathaiah}

\begin{figure}
\includegraphics[scale=0.30,angle=-90]{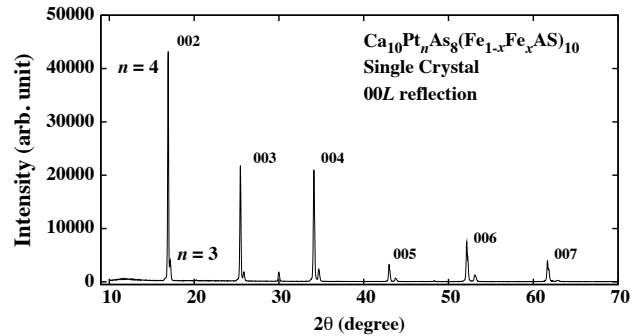}
\caption{\label{fig:diff}
X-ray diffraction intensity of $00L$ reflections of a part of the crystal used for the neutron scattering measurements.
}
\end{figure}

A single crystal, which weighs $\sim 10$ g, was prepared from the starting mixture of CaAs : FeAs : Pt : Fe with the molar ratios 1.00 : 1.00 : 0.51 : 0.09 (Ca : Fe : Pt : As = 1.84 : 2.00 : 0.94 : 3.67).
The susceptibility measurement on a part of the crystal with a SQUID magnetometer exhibits a significant shielding diamagnetism below $T_{\rm c}^{\rm onset} \sim$ 33 K as shown in Fig.~\ref{fig:xtal}(b).
The distance between the FeAs layers (or lattice parameter $c$) of the pseudo tetragonal unit cell estimated from the $00L$ reflections (Fig.~\ref{fig:diff}) is $\sim$ 10.51 \AA.
In the same figure, we can clearly distinguish the epitaxially grown second phase Ca$_{10}$Pt$_3$As$_8$(Fe$_{1-x}$Pt$_x$As)$_{10}$ with the FeAs-layer distance $c$ of $\sim$ 10.33 \AA.
This difference between the $c$ values of these two kinds of phases is commonly found in many papers published previously.
\begin{figure*}
\includegraphics[scale=0.60,angle=-90]{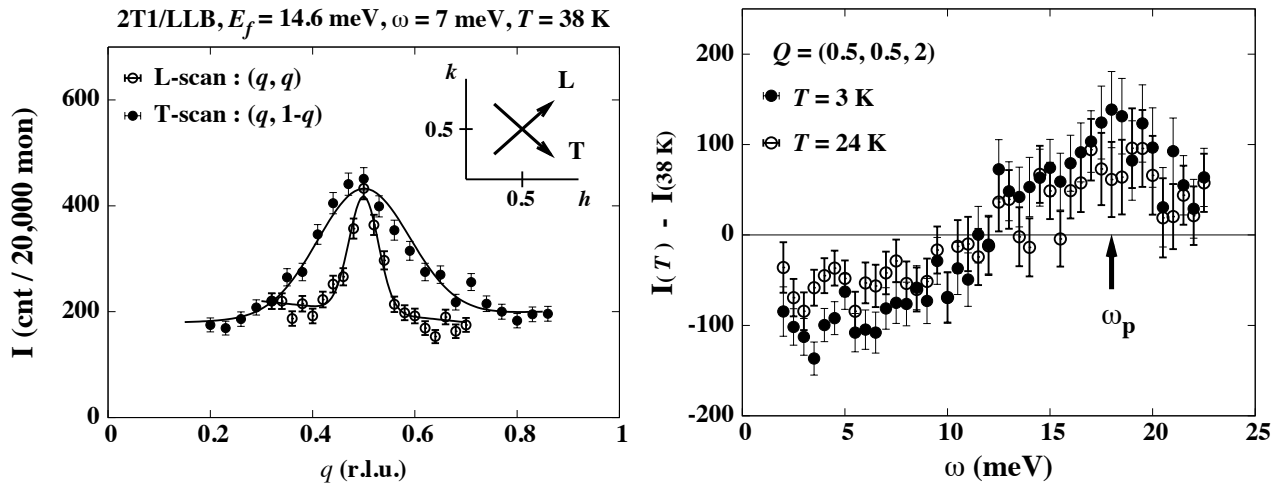}
\caption{\label{fig:mag}
(Left) ${\bm Q}$ scan profiles of the magnetic excitation spectra measured along L- and T-directions (inset), respectively, through the M point ${\bm Q}$ = (0.5, 0.5, 0) at $T$ = 38 K.
(Right) The difference of the intensities of the magnetic excitations at the M point ${\bm Q}$ = (0.5, 0.5, 2) between $T$ = 3 K (SC) and 38 K (normal) (solid circles), and between 24 K (SC) and 38 K (normal) (open circles).
The arrow indicates the energy of the peak position of the difference.
}
\end{figure*}
Judging from the very small fraction of the Bragg reflection intensities of the second phase, we can find that the second phase does not affect results of the present studies on their dynamical properties.
As shown in the Appendix, the $x$ value is estimated to be 0.05--0.06, which is smaller than the value reported previously.\cite{22sato3,23ikeuchi,24sato4,26kobayashi}
The rapid growth of the shielding diamagnetism with decreasing $T$ through $T_{\rm c}^{\rm onset}$, the SC transition occurs in the major part of the sample just below $T_{\rm c}^{\rm onset}$, or at least, within the $T$ range of a few degree below $T_{\rm c}^{\rm onset}$.
We will return this point later in relation to the magnetic excitation data.
In the studies of the NMR Knight shift　and $1/T_1$, X-ray structural analyses, magnetic susceptibility and electrical resistivity, no evidence for structural and antiferromagnetic transitions has been observed,\cite{25ni2,26kobayashi,39sturzer2} which is contrasted to the case of Ca$_{10}$Pt$_3$As$_8$(Fe$_{1-x}$Pt$_x$As)$_{10}$.\cite{29sturzer,30sapkota,31gao}

 Data of the spin excitations and low-energy phonons were collected with both the thermal (2T1) and cold triple-axis-spectrometers (4F2) at the neutron reactor ORPHEE of Laboratoire Leon Brillouin (LLB), France, respectively, where incident and scattered beams were focused by pyrolytic graphite 002 monochromater and analyzer, respectively.
The collimation conditions were fully open. Neutron data of optical phonons were collected with the thermal triple-axis spectrometer (HB-3) at HFIR of Oak Ridge National Laboratory, USA. The collimation condition was 48$'$-40$'$-40$'$-70$'$.
In all cases, pyrolytic graphite filters were placed before the analyzer to eliminate the higher order reflections.

The phonon energies have been calculated for Ca$_{10}$Pt$_4$As$_8$(FeAs)$_{10}$ by the first-principle calculation package, Vienna Ab initio Simulation Package(VASP),\cite{32kresse} where the structure of Ca$_{10}$Pt$_4$As$_8$(FeAs)$_{10}$ optimized before the phonon calculation was used (The crystal parameters thus obtained are in the previous report.\cite{33machida}; space group $P4/n$).
The calculated phonon results are compared with the experimentally observed ones.

\section{results and discussion}

\subsection{\label{sec:level2}Magnetic scattering}
First, we briefly show results of magnetic scattering obtained for the crystal sample of Ca$_{10}$Pt$_4$As$_8$(Fe$_{1-x}$Pt$_x$As)$_{10}$.
The typical ${\bm Q}$-scan profiles of the magnetic scattering intensity $I_{\rm mag}({\bm Q}, \omega) = (n+1)\,\chi ''_{\rm mag}({\bm Q}, \omega)$ obtained at constant-$\omega$ with the scattered neutron energy $E_f$ = 14.6 meV are shown in the left panel of Fig.~\ref{fig:mag}, $n$ being the Bose factor, where the directions of the longitudinal and transverse scans through the M point (1/2, 1/2, 0) (L- and T-scans, respectively) are also shown in the inset.
The anisotropy of the profile width is clearly observed as reported commonly for various other Fe-based systems.\cite{036lester,037diallo,038li}
The difference of the magnetic scattering intensities between 38 K ($> T_{\rm c}$) and two temperatures, 24 K and 3 K ($< T_{\rm c}$) observed at another M point ${\bm Q}$ = (1/2, 1/2, 2) are in the right panel of Fig.~\ref{fig:mag}.
From the figure, we can see the following.
(1) With decreasing $T$ through $T_{\rm c}$, or with increasing SC order parameter $\Delta$, the magnetic scattering intensity in the low-energy region is suppressed and it is enhanced in the high-energy region, and the maximum of the enhancement is found at $\omega = \omega_{\rm p} \sim$ 18 meV, which corresponds to $\omega_{\rm p}/k_{\rm B}T_{\rm c}^{\rm onset} \sim$ 6.3.
(2) The spectral weight was found to be zero in the low $\omega$-region (at least $\omega \lesssim 7$ meV at 3 K), indicating that the SC state is established above 24 K all over the whole volume, as stated in section II.
The peak of the enhancement, is usually expected for SC order parameters $\Delta$ with the sign reversal, and called ``the resonance peak".
For Fe-based systems, as Maier and Scalapino\cite{039maier} pointed out, it appears, if the spin-fluctuation mechanism is relevant to the superconductivity, at $\omega_{\rm p}$ smaller than the value of $|\Delta_{\Gamma}|+|\Delta_{\rm M}|$, $\Delta_{\Gamma}$ and $\Delta_{\rm M}$ being the order parameters at the Fermi surfaces around the $\Gamma$ and M points, respectively.
However, it was also pointed out in Refs.~\onlinecite{08onari2} and \onlinecite{11onari3} that even for the $S_{++}$ symmetry of $\Delta$, such a peak can be expected at $\omega_{\rm p} \geq |\Delta_{\Gamma}|+|\Delta_{\rm M}|$, if we consider the electron dissipation effect.
On this point, we cite Ref.~\onlinecite{34surmach}, which shows that the observed relation $\omega_{\rm p}/k_{\rm B}T_{\rm c} \sim$ 4.3 holds for various systems, and that larger values of $\omega_{\rm p}/k_{\rm B}T_{\rm c}$ have been found for Ca$_{10}$Pt$_n$As$_8$(Fe$_{1-x}$Pt$_x$As)$_{10}$ ($n$ = 3 and 4) systems, including the present data $\omega_{\rm p}/k_{\rm B}T_{\rm c}^{\rm onset} \sim$ 6.3.
Because it is not easy to experimentally determine the $\Delta$ values accurately, we cannot distinguish which one of the $S_{\pm}$ and $S_{++}$, the relation $\omega_{\rm p}/k_{\rm B}T_{\rm c}^{\rm onset} \sim$ 4.3 itself supports.
Then, the relation $\omega_{\rm p}/k_{\rm B}T_{\rm c}^{\rm onset} \sim$ 6.3 observed here, which is favorable to the $S_{++}$ symmetry, is rather encouraging.
(Note that $\omega_{\rm p}/k_{\rm B}T_{\rm c}^{\rm onset}$ observed here is regarded as its minimum estimation in the sense that we used the onset value $T_{\rm c}^{\rm onset}$ instead of $T_{\rm c}$.) 
Then, we proceeded the phonon measurements, the results of which are described in the next section.

\subsection{\label{sec:citeref}Phonons}
\begin{figure}
\includegraphics[scale=0.65]{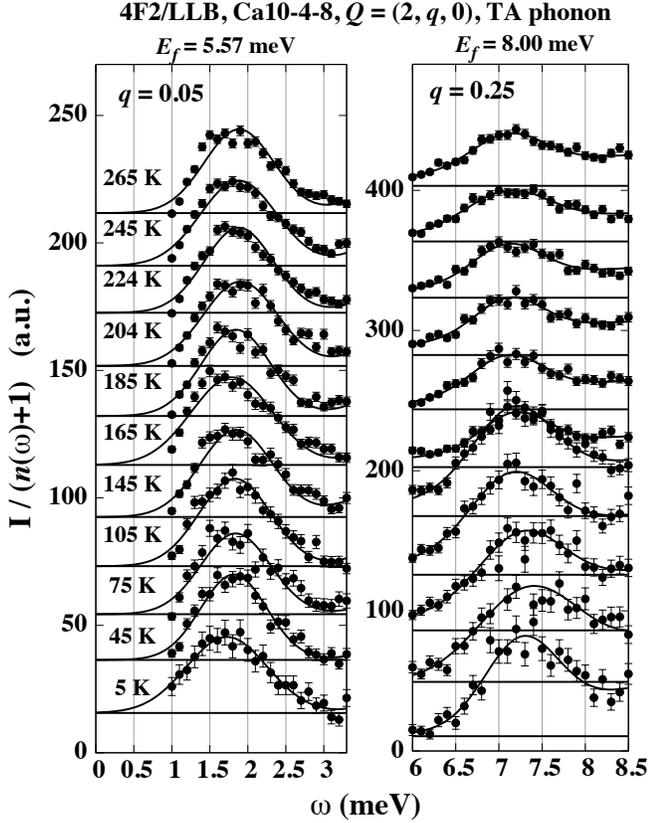}
\caption{\label{fig:T200}
The TA phonon profiles obtained by scanning the transfer energy $\omega$ at ${\bm Q} = (2, 0, 0)+{\bm q}$ with two ${\bm q}$ vectors of (0, 0.05, 0) and (0, 0.25, 0) are shown.
The solid curves are fitted lines.
The horizontal lines are their baseline, shifted arbitrarily for each temperature from 5 K to 260 K.
}
\end{figure}

\begin{figure}
\includegraphics[scale=0.39,angle=-90]{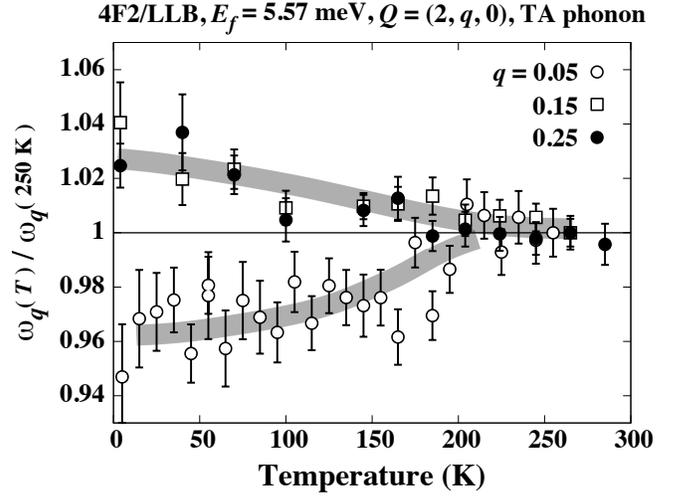}
\caption{\label{fig:softening}
Temperature dependences of the phonon frequencies of the TA branch measured at ${\bm Q} = (2, 0, 0)+{\bm q}$ are shown for ${\bm q} = (0, q, 0), (q = 0.05, 0.15,$ and $ 0.25)$ in the form of $\omega_q(T)/\omega_q(T = 250$ K)$-T$ plot.
The curves in the figure are guides for the eye.
}
\end{figure}

Phonon measurements have been carried out for two kinds of modes with main characters of the in-plane transverse acoustic (TA) phonon near the $\Gamma$ point and in-plane optical modes at an M point mainly polarized within the FeAs plane, because the orbital fluctuation is theoretically expected to be strong at these points.\cite{9kontani1,13kontani2,14inoue}
The former phonon corresponds to the elastic constant $C_{66}$, which is well-known to exhibit the softening in Ba(Fe$_{1-x}$Co$_x$)$_2$As$_2$ as $T$ decreases toward the tetragonal-orthorhombic transition point $T_{\rm S}$.\cite{041goto,042yoshizawa}
In the present case, the corresponding TA phonon was measured at several ${\bm Q} = (2, 0, 0)+{\bm q}$ and $T$ points with ${\bm q} = (0, q, 0)$.
The observed $\omega$-scan profiles are shown in Fig.~\ref{fig:T200} at two fixed $q$-values of 0.05 and 0.25, for example.
They were obtained with the neutron energy-loss condition and with the final energy being fixed at $E_f = 5.57$ meV and $E_f = 8.00$ meV, respectively.
Before the measurement at $q$ = 0.05, the Bragg point (2, 0, 0) was adjusted at each $T$ point to ensure an accurate value of $q$.
The phonon energies were determined by fitting the Gaussian line shape as shown in Fig.~\ref{fig:T200}, where the intensity is divided by the Bose population factor $n+1$.
The results are shown in Fig.~\ref{fig:softening} in the form of $\omega_q(T)/\omega_q(T = 250$ K) for $q$ = 0.05, 0.15 and 0.25.
(Calculations using the convolution of the resolution function gave essentially the same results.)
In the figure, we can see the softening tendency of the phonon energy at $q$ = 0.05 as compared to those at the higher $q$ points with decreasing $T$, that is, the energy at $q$ = 0.05 shifts to the lower-energy side, while those at $q$ = 0.25 have the opposite tendency.
The relative magnitude of the softening fraction extrapolated to ($q$, $T$)$\rightarrow$0 is consistent with the data of the elastic constant $C_{66}$ reported for the optimal or slightly overdoped Ba(Fe$_{1-x}$Co$_x$)$_2$As$_2$,\cite{042yoshizawa} which is also SC and exhibit none of the magnetic and structural transitions.
(Here, note that $\omega \propto \sqrt{C_{66}}$.) 

The dispersion curves of this in-plane TA branch was compared with those obtained by the first principle calculation for the pseudo tetragonal cell, where the calculated and experimental results were found to agree reasonably well , even though we used the crystal structure of Ca$_{10}$Pt$_4$As$_8$(FeAs)$_{10}$ optimized by the calculation.


\begin{figure}
\includegraphics[scale=0.37,angle=-90]{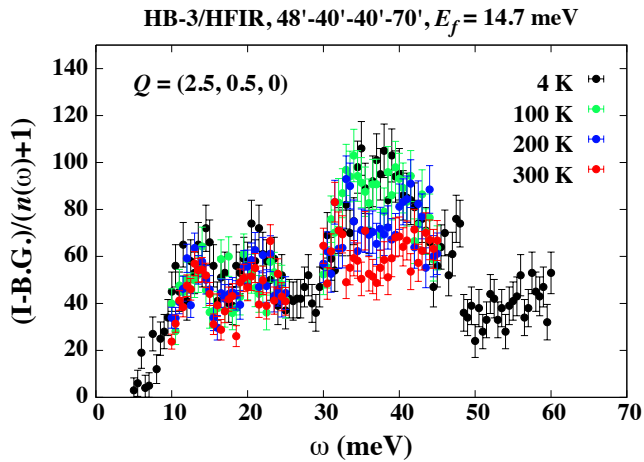}
\caption{\label{fig:optic}
Phonon spectral weight at the M point ${\bm Q}$ = (2.5, 0.5, 0) is shown for several $T$ points.
These spectra are corrected by Bose factor after subtracting the $\omega$-independent background determined in the high energy region.
}
\end{figure}

We next move to the study of the in-plane optical phonons.
The phonon intensity was measured at the fixed scattering vector ${\bm Q}$ = (2.5, 0.5, 0)  by scanning the energy transfer $\omega$, where the final neutron energy was fixed $E_f$ = 14.7 meV.
Note that this spectrometer setting is to see the phonon modes mainly having in-plane polarizations at the M point.
Then, after subtracting the $\omega$-independent background intensity (B.G.) determined in the high energy region from the raw data $I_{\rm ph}({\bm Q}, \omega)$, we have obtained the phonon spectral weight $\chi ''_{\rm ph}({\bm Q}_{\rm M}, \omega) = [I_{\rm ph}({\bm Q}_{\rm M}, \omega)-{\rm B. G.}] /(n+1)$, and the results are shown in Fig.~\ref{fig:optic} against $\omega$ for various $T$ values.
We can find significant $T$ dependence of $\chi ''_{\rm ph}({\bm Q}_{\rm M}, \omega)$ in the region of $\omega$ around 37 meV.
Note that no spectral weight exists above 40 meV, as shown below by the calculation.
The data for $\omega >$ 45 meV are just due to the excess B.G. at ${\bm Q}$ = (2.5, 0.5, 0).

\begin{figure}
\includegraphics[scale=0.57]{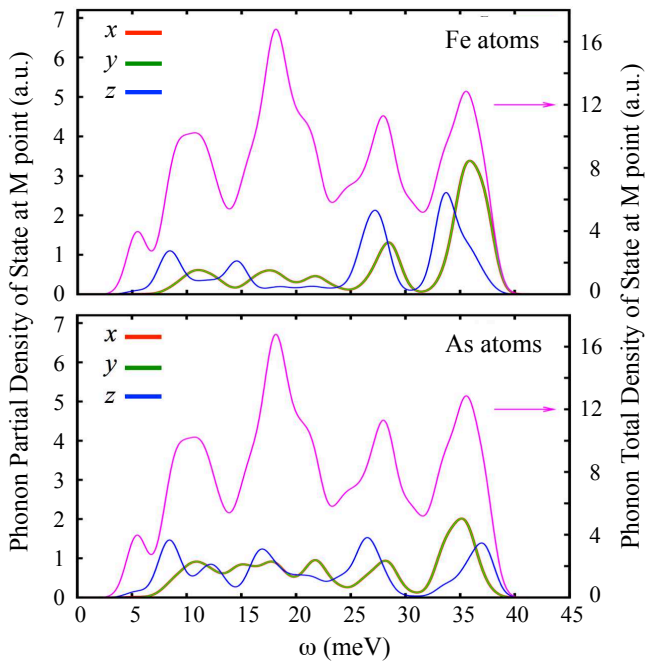}
\caption{\label{fig:calc}
Calculated phonon density of state defined as described in the text is shown at M point for (top) Fe atoms and (bottom) in-plane As atoms.
Each component of the vibrational direction can be distinguished by the colors of the line.
The total density of states represents the sum of the ones with $i = x$, $y$, and $z$.
Note that the red and green lines almost overlap with each other.
}
\end{figure}

To study the origin of this strong $T$ dependence, we first calculated, by the first principle calculation,\cite{33machida} the partial phonon density of states 
$g^{(i)}({\bm Q}, \omega) = \sum_j\,{\bm e}^{(i)}_j({\bm Q})\ \delta(\omega-\omega_j({\bm Q}))$ with ${\bm e}^{(i)}_j({\bm Q})$ being the $i$ ($x$, $y$, or $z$) component of the polarization vector of the $j$-th mode (or we often specify the atomic element of the $j$-th mode, too, by only this $j$).
The results are shown for the $i$-component of Fe and As motions in Figs.~\ref{fig:calc}(a) and \ref{fig:calc}(b), respectively, where we find that the peak positions of the total density of states seem to be slightly shifted to the lower side from the observed peak positions at 300 K.
We can also see that all modes at the M point are in the region less than 40 meV and the peak at the largest energy originates mainly from the in-plane vibrations of Fe and As atoms.
Although the observed phonon intensity (Fig.~\ref{fig:optic}) cannot be compared directly with the calculated data shown in Fig.~\ref{fig:calc}, partly because the former has the factor $|\,{\bm e}^{(i)}_j({\bm Q})\cdot {\bm Q}\,|^2$ instead of ${\bm e}^{(i)}_j({\bm Q})$ in the latter, we can, at least, say that the observed  highest-energy peak of $g^{(i)}({\bm Q}, \omega)$ corresponds the in-plane motions of FeAs layers.

\begin{figure}
\includegraphics[scale=0.38,angle=-90]{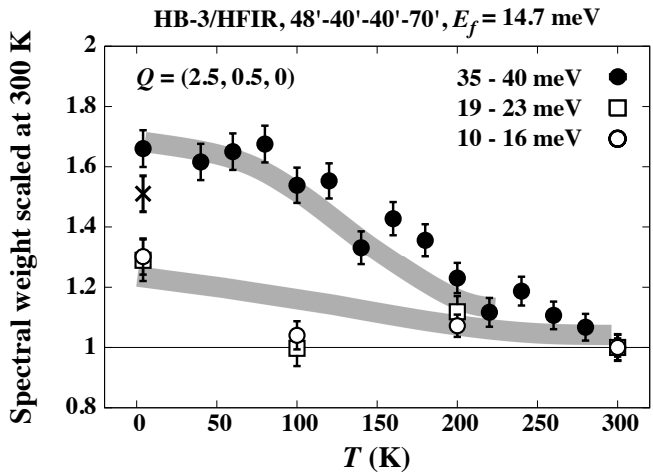}
\caption{\label{fig:weight}
$T$ dependences of spectral weights integrated over the regions 10--16, 19--23 and 35--40 meV are shown by open circles, open squares, and closed circles, respectively.
Cross symbols show results of 35--40 meV region taken at ${\bm Q}$ = (2.50, 0.25, 0).
Data are scaled by the values at $T$ = 300 K for each ${\bm Q}$ point.
The curves are the guides for the eye.
}
\end{figure}

From this view point, we plot in Fig.~\ref{fig:weight} the observed phonon intensity integrated over the $\omega$ region of 35--40 meV against $T$, together with those in the regions of 10--16 and 19--23 meV, where the difference among the $T$ dependences of $\chi ''_{\rm ph}({\bm Q}_{\rm M}, \omega)$ are explicitly seen.
The data at ${\bm Q}$ = (2.5, 0.25, 0) are also shown by crosses taken at 5 K and room temperature.


Here, we add the following point.
First, the anomalous $T$ dependence of the phonon intensity cannot be explained by considering the so-called ``false peaks'', which may appear accidentally due to the spectrometer condition (see Ref.~\onlinecite{35shirane} for details).
Second, effects of the Debye-Waller factor on the $T$ dependence of phonon intensities is almost independent of the phonon energy $\omega$\cite{36kittel} at a fixed ${\bm Q}$ point, which is also confirmed by the first principle calculation in the present case at ${\bm Q_{\rm M}} = (2.5, 0.5, 0)$.
The calculation has also confirmed that the Debye-Waller effect is much smaller than the observation at around $\omega \sim$ 37 meV, but seems to be consistent with those in the other $\omega$ regions.

Now, we have found the gradual but rather significant increase of the phonon spectral weight of the in-plane optical modes.
It starts at $\sim$250 K with decreasing $T$, and is reminiscent of the similar $T$ dependence of various static quantities observed in Fe-based systems in the region between the structural transition ($T_{\rm S}$) and nematic ($T^*$) temperatures.\cite{18chu,19yi,20shimojima,045miao,046kasahara}
(Note that $T_{\rm S} = 0$ in the present sample.)
In this $T$ region, the breakdown of the four-fold symmetry, in particular, the splitting of the $3d_{yz}$ and $3d_{zx}$ orbitals have been found\cite{19yi,20shimojima} suggesting the contribution of the strong orbital fluctuation to stabilize the local lattice distortion.
Then, the observed softening of the in-plane TA phonon and the unusual $T$ dependence of the in-plane optical phonons should contain information how the lattice system couples to the orbitals (and/or spins).
In turn, it can be used to distinguish which one of ``spins or orbitals'' is primarily important for the realization of the nematic phase and superconductivity.

It is worth noting that if phonons primarily contribute to the pairing mechanism, unusual $T$ dependence of $\chi ''_{\rm ph}({\bm Q}, \omega)$ is expected as the superconductivity occurs at the phonon energy $\omega \sim 2\Delta$,\cite{38maier} which is much smaller than the energy region of the observed unusual $T$ dependence (around $\sim$ 37 meV).
This excludes the possibility of the primary relevance of the phonons to the SC mechanism.
(Note that the orbital fluctuation mechanism itself does not require the primary role of phonons.)
Finally, the strong $T$ dependence of $\chi ''_{\rm ph}({\bm Q}, \omega)$ does not seem to be confined within the ${\bm Q}$ region very close to the M point, as shown by the data at ${\bm Q}$ = (2.5, 0.25, 0), which remains as a future issue to be clarified by detailed studies.

\section{summary}
We have carried out inelastic neutron scattering measurements of phonons as well as magnetic excitations of a single crystal of Ca$_{10}$Pt$_4$As$_8$(Fe$_{1-x}$Pt$_x$As)$_{10}$.
The maximum of the $\chi''_{\rm mag}({\bm Q}_{\rm M}, \omega)$-enhancement by the superconductivity is observed at $\omega = \omega_p \sim$ 18 meV, which corresponds to $\omega_{\rm p}/k_{\rm B}T_{\rm c}^{\rm onset} \sim$ 6.3, rather favorable to the $S_{++}$ symmetry of the order parameter expected by the orbital fluctuation mechanism.
In the measurements of the in-plane TA phonons, the softening at the $\Gamma$ point has been observed, suggesting that it is a common phenomenon of Fe-based superconductors.
In the measurement of the in-plane optical phonons at the M-point ${\bm Q}$ = (2.5, 0.5, 0), the strong $T$ dependence of the phonon spectral weight $\chi''_{\rm ph}({\bm Q}_{\rm M}, \omega)$ was found.
These phonon behaviors become appreciable with decreasing $T$ far above $T_{\rm c}$.
The results obtained here present information to understand what roles the orbital fluctuation plays in realizing the superconductivity and also nematic behavior found in various Fe-based systems.

\begin{acknowledgments}
Authors thank Prof. H. Kontani for fruitful discussion.
The work is supported by Grants-in-Aid for Scientific Research (Grant No. 24340080) from the Japan Society for the Promotion of Science (JSPS) and Technology and JST, TRIP.
The experiments at ORPHEE at Saclay was carried out by a project No.10604.
The measurements at High Flux Isotope Reactor (HFIR) at Oak Ridge National Laboratory (ORNL) were carried out by a project No. ITPS-7059 with the supports of the travelling expense by U.S.–Japan Cooperative Program on Neutron Scattering Research.
Research at ORNL's HFIR was sponsored by the Division of Scientific User Facilities of the Office of Basic Energy Sciences, US Department of Energy.
\end{acknowledgments}

\appendix
\section{Pt content in the FeAs planes}
\renewcommand{\figurename}{FIG. A}
\setcounter{figure}{0}
In our previous reports, the $x$ value of the present sample was estimated to be $\sim$ 0.2 by analyzing the line shapes of $^{75}$As NQR and NMR-centerline spectra.\cite{26kobayashi}

\begin{figure}
\includegraphics[scale=0.65,angle=-90]{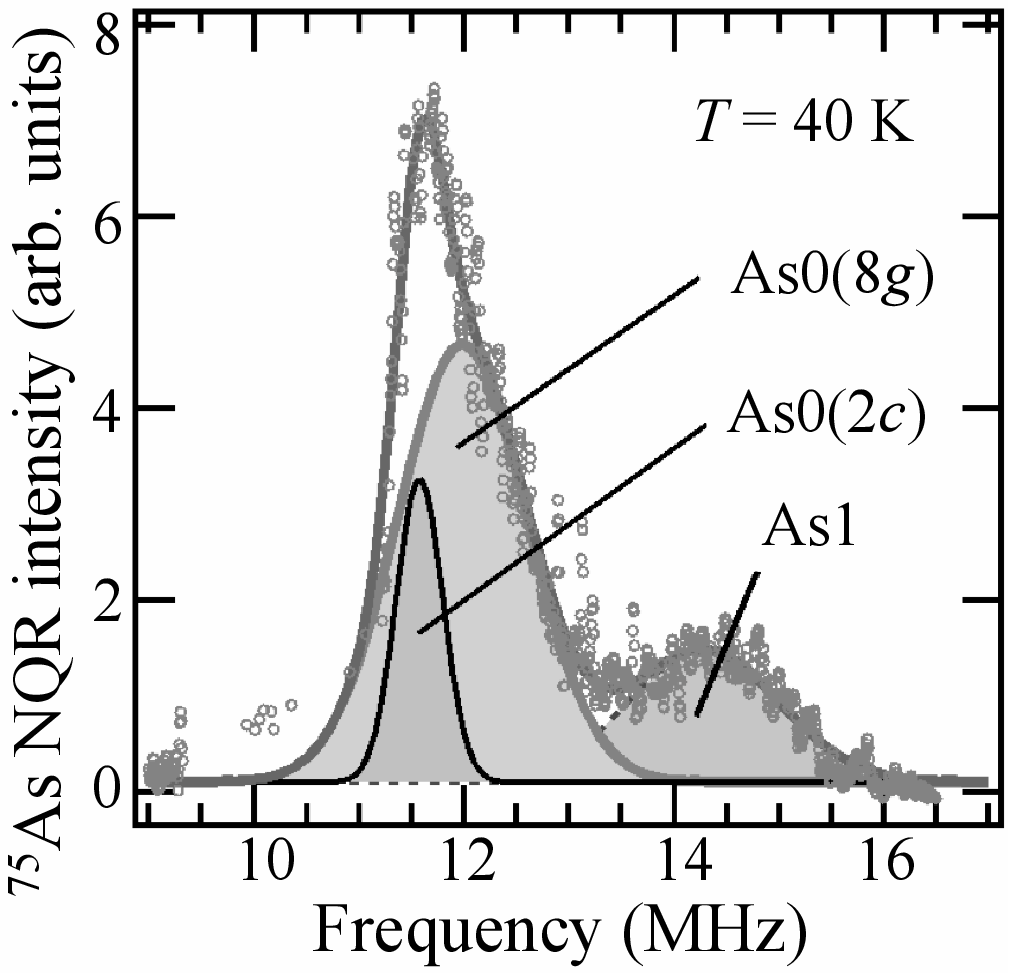}\\[3mm]
\includegraphics[scale=0.8,angle=0]{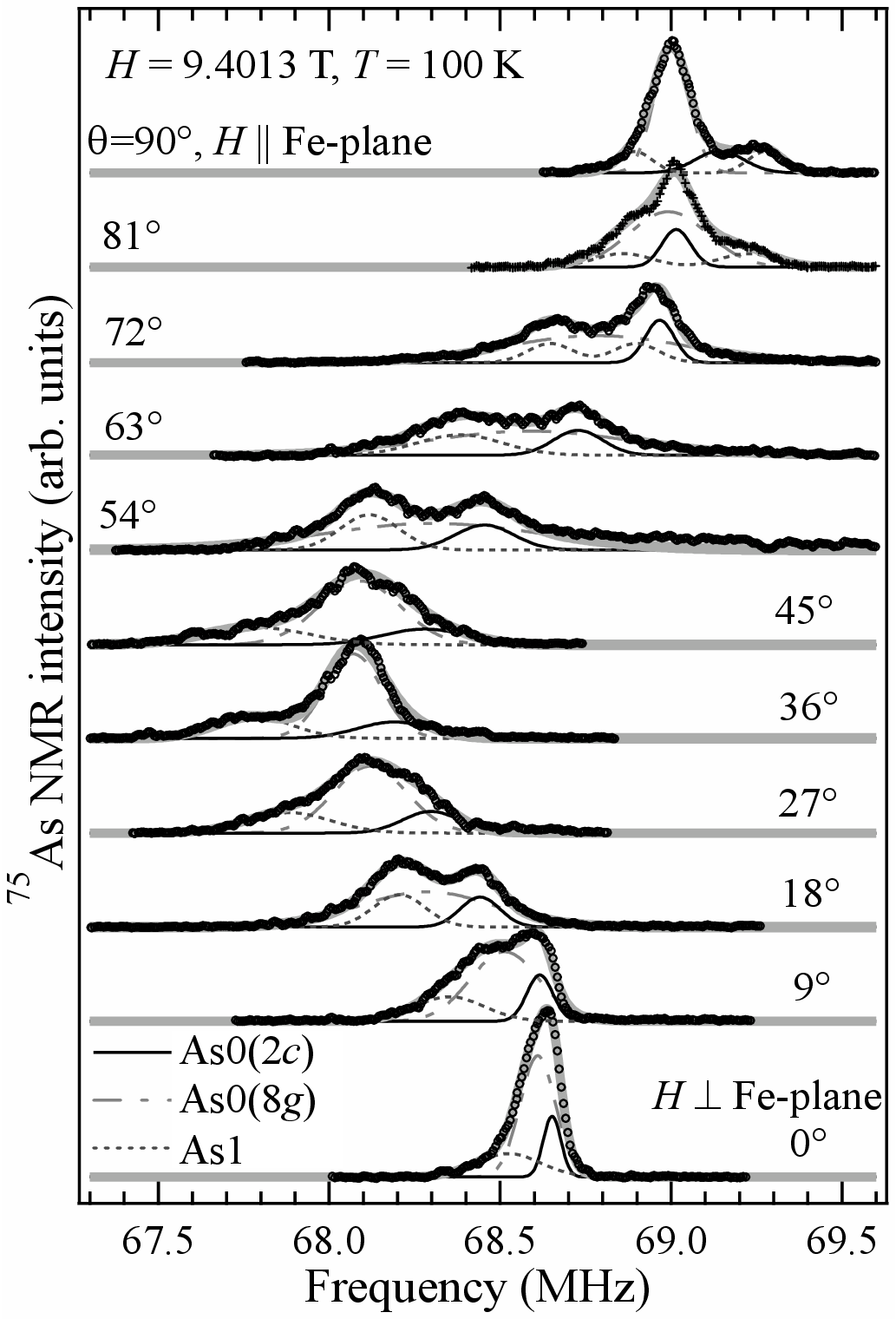}
\caption{\label{fig:a1}
The NQR profiles at 40 K (top) and angular ($\theta$) dependence of the NMR profiles (bottom) taken at 100 K with the applied magnetic field $H = 9.4013$ T are shown with the fitted curves (solid lines) obtained for $x = 0.06$, $\theta$ being the angle between ${\bm H}$ and the $c$ axis of the pseudo tetragonal cell.
The $N$ value in As$N$ indicates the Pt number doped to the four nearest neighbor Fe sites, and 2$c$ and 8$g$ in the parentheses indicate the two crystallographically distinct As0 sites (space group: $P4/n$).
In the calculation, the ratios of the signal intensities of As0(2$c$), As0(8$g$) and As1 sites were approximated as 0.16 : 0.62 : 0.22 by considering the binominal distribution of Pt impurities doped to Fe sites of FeAs planes.\cite{26kobayashi}
(Note that we did not consider the difference between the two distinct sites of As1(2$c$) and As1(8$g$) , because it does not result in meaningful changes.)
}
\end{figure}

However, considering the data reported in Refs.~\onlinecite{25ni2}, \onlinecite{39sturzer2}, and \onlinecite{40lohnert}, we have checked the $x$ value again, and found that, by using the value $x = 0.06$, the NQR and NMR data can be explained better than the previous case, as shown in Fig. A~\ref{fig:a1}.
Because it is consistent not only with the existing reports,\cite{25ni2,39sturzer2,40lohnert} but also with the used starting composition, we revise the value.
By this revision, no essential qualitative changes of the results in the previous reports emerge except the magnitudes of the NMR Knight shift shown in Fig.~\ref{fig:mag} of Ref.~\onlinecite{26kobayashi}.


\bibliography{CaPtFeAs_phonon}

\end{document}